%% file: root.tex
\documentclass{ifacconf}
\usepackage{eso-pic}

\usepackage[utf8]{inputenc}
\usepackage[T1]{fontenc}
\usepackage{breqn}
\usepackage{amsmath}
\usepackage{amsfonts}
\usepackage{amssymb}
\newtheorem{remark}{Remark}

    \usepackage{enumitem}
\usepackage{tikz}
\usetikzlibrary{shapes,arrows,calc,fit}
\tikzset{
        block/.style = {draw, rectangle,
            minimum height=1cm,
            minimum width=2cm},
        input/.style = {coordinate,node distance=1cm},
        output/.style = {coordinate,node distance=4cm},
        arrow/.style={draw, -latex,node distance=2cm},
        pinstyle/.style = {pin edge={latex-, black,node distance=2cm}},
        sum/.style = {draw, circle, node distance=1cm},
    }
\usepackage{pgfplots} 
\pgfplotsset{compat=newest} 
\pgfplotsset{plot coordinates/math parser=false} 
\newlength\figureheight 
\newlength\figurewidth 
\pgfplotsset{
    every axis plot post/.style={
        line join=round
    }
}
\input{tikz_set}

\usetikzlibrary{external} 
\usepackage{graphicx}      
\usepackage{natbib}        

\AddToShipoutPictureBG*{%
	\AtPageUpperLeft{%
		\setlength\unitlength{1in}%
		\hspace*{\dimexpr0.5\paperwidth\relax}
		\makebox(0,-2)[c]{\begin{tabular}{c c}
				Max van Meer, Optimal Commutation for Switched Reluctance Motors using Gaussian Process Regression, \\
				To appear in {\em 2022 IFAC Modeling, Estimation and Control Conference}, Jersey City, New Jersey, 2022,\\uploaded to ArXiv \today
		\end{tabular}}
}}

\begin{document}
\begin{frontmatter}

\title{Optimal Commutation for\\ 
Switched Reluctance Motors\\
using Gaussian Process Regression} 

\thanks[footnoteinfo]{This work is part of the research programme VIDI with project number 15698, which is (partly) financed by the Netherlands Organisation for Scientific Research (NWO). In addition, this research has received funding from the ECSEL Joint Undertaking under grant agreement 101007311 (IMOCO4.E). The Joint Undertaking receives support from the European Union’s Horizon 2020 research and innovation programme.}

\author[TUE]{Max van Meer} 
\author[TUE,TNO]{Gert Witvoet} 
\author[TUE,Delft]{Tom Oomen}

\address[TUE]{Eindhoven University of Technology, 
   Eindhoven, the Netherlands (e-mail: m.v.meer@tue.nl).}
   \address[TNO]{TNO, Dept. of Optomechatronics, Delft, the Netherlands.}
   \address[Delft]{Delft University of Technology, Delft, the Netherlands.}
\begin{abstract}                
Switched reluctance motors are appealing because they are inexpensive in both construction and maintenance. The aim of this paper is to develop a commutation function that linearizes the nonlinear motor dynamics in such a way that the torque ripple is reduced. To this end, a convex optimization problem is posed that directly penalizes torque ripple in between samples, as well as power consumption, and Gaussian Process regression is used to obtain a continuous commutation function. The resulting function is fundamentally different from conventional commutation functions, and closed-loop simulations show significant reduction of the error. The results offer a new perspective on suitable commutation functions for accurate control of reluctance motors.
\end{abstract}


\end{frontmatter}

\section{Introduction}
\begin{figure}[tb]
\centering
\includegraphics[width=0.6\linewidth]{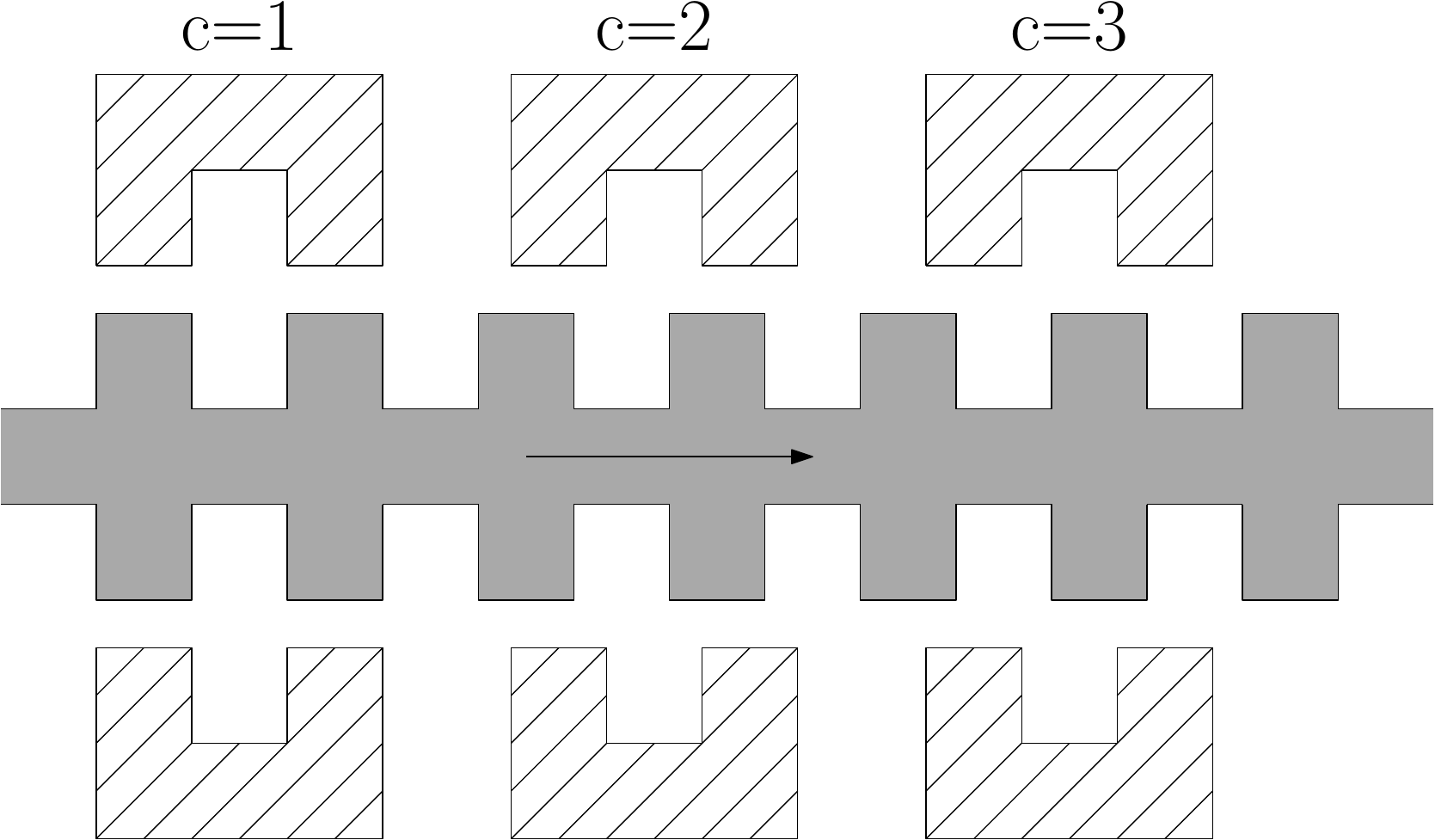}
\vspace*{-3mm}\caption{Schematic overview of the Switched Reluctance Motor. By sequentially  applying current waveforms to the coils, a torque is applied to the toothed rotor.}
\label{fig:SRM}
\end{figure}

\begin{figure}[tb]
\centering
\includegraphics[width=1\linewidth]{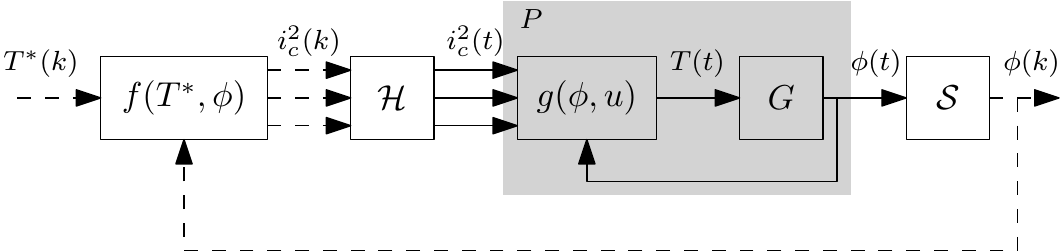}
\vspace*{-3mm}\caption{Overview of the commutation strategy. 
}
\label{fig:controlscheme}
\end{figure}

Switched Reluctance Motors (SRMs) offer numerous advantages over DC or AC drives due to their mechanical simplicity, efficiency, low cost, and robust construction \citep{Miller1993,Katalenic2013}, see Figure \ref{fig:SRM}. Its application is particularly attractive when maintenance is expensive or the costs should be low, e.g., in optical communication terminals on satellites \citep{Kramer2020}. On the other hand, a main drawback of SRMs is the complexity in controlling its position, due to the nonlinear relationship between torque, rotor angle, and currents to the different phases. These nonlinear dynamics are typically attempted to be linearized using a commutation strategy \citep{Wang2016}, as shown in Figure \ref{fig:controlscheme}, and if done imperfectly, this will introduce a torque ripple to the rotor.

The problem of finding a commutation strategy is inherently over-parametrized, since at any rotor position, multiple coils can apply a torque to the rotor. Model-based methods of designing a commutation strategy assume that a description of the nonlinear relation between torque, current and rotor angle is available, and define a torque sharing function to divide the currents over the coils \citep{Balaji2004,Ilic-Spong1987}. The specification of this torque sharing function is often done in an ad-hoc manner \citep{Wang2016,Vujicic2012}, and experimental validation is required to determine which function leads to the least torque ripple, since each function is a valid solution of the over-parametrized commutation problem.


In a different line of developments, non-parametric modeling has attracted increased attention in the last decades due to its flexible model structures. In particular, Gaussian Process (GP) regression \citep{Rasmussen2004} has been applied extensively to motion control \citep{Poot2022}, since it admits intuitive model structures based on properties such as smoothness and periodicity, and allows for automatic optimization of hyper-parameters to fit a continuous function through a finite number of points. 


Although conventional model-based design methods for commutation strategies lead to a highly flexible control scheme that allows for the use of linear controllers, current design techniques of commutation functions do not take a major cause of torque ripple into account. In fact, this paper shows that even if a perfect model is available, torque ripple still occurs because of sampling effects. If the sampling rate is decreased or the velocity is increased, this sampling-induced torque ripple increases. Indeed, low-cost applications that would benefit from the simplicity of SRMs may not allow for high sampling rates.

Therefore, the aim of this paper is to reduce sampling-induced torque ripple in the presence of a perfect model, by formulating a commutation function that divides currents between coils specifically to reduce sampling-induced torque ripple. To this end, an optimization problem is posed that penalizes torque ripple and power consumption as a function of the commutation function. The problem is solved to find realizations of an optimal commutation function at a finite number of motor angles. Subsequently, a prior is posed on the commutation function, based on smoothness and periodicity, and Gaussian Process regression is used to obtain a continuous commutation function. This leads to the following contributions: \begin{enumerate}[label={C\arabic*:}]
\item A design method for commutation functions of switched reluctance motors is developed that allows for an intuitive design trade-off between power consumption and sampling-induced torque ripple.
\item The effectiveness of the commutation strategy is verified in simulation, and shown to lead to commutation functions that vary significantly from conventional commutation functions. 
\end{enumerate}
This paper is structured as follows. In Section \ref{sec:problem}, the setting and commutation strategy are given, and it is explained how sampling can introduce torque ripple. Subsequently, Section \ref{sec:optimal} describes the developed design method for a commutation strategy that reduces sampling-induced torque ripple. Next, Section \ref{sec:sim} describes the simulation results, and finally, \ref{sec:conclusion} presents the conclusion, as well as directions for further research. 






\section{Problem description}\label{sec:problem}
In this section, the problem description is given. First, the working principle of a switched reluctance motor is explained, which serves as the motivating example for the research in this paper. Next, the system and the commutation strategy are described. Subsequently, it is shown that torque ripple appears even in the presence of a perfect model, because of sampling effects. Finally, the problem formulation is given.

\subsection{Switched reluctance motor concept}
The SRM, depicted schematically in Figure \ref{fig:SRM}, consists of a rotor with $n_{\text{t}}$ teeth, positioned between three coils \citep{Kramer2020}. Each coil $c\in[1,2,3]$ can be magnetized separately by application of a current $i_c(t)\in\mathbb{R}$ at time $t\in\mathbb{R}$. Neglecting magnetic saturation, the resulting torque on the rotor is given \citep{Miller1993} by \begin{equation}\label{eq:reluctance}
T_{c}\left(\phi, i_{c}\right)=\frac{1}{2} \frac{d L_{c}(\phi)}{d \phi} i_{c}^{2},
\end{equation}
where $L_c(\phi)$ is the phase inductance as a function of rotor position $\phi$. The inductance is maximal when the rotor teeth are aligned with the teeth that are magnetized by the coil, and minimal when these do not overlap. Hence, $L_c(\phi)$ is periodic in $\phi$ with spatial period $\frac{2\pi}{n_{\text{t}}}$. Since the required current to generate a torque depends on $\phi$ in \eqref{eq:reluctance}, a commutation strategy is desired that divides currents between the coils based on $\phi$ to achieve some desired torque. This is explained in detail in the next section. 
\subsection{Setting and goal}\label{sec:setting}
\begin{figure}[tb]
\centering
\input{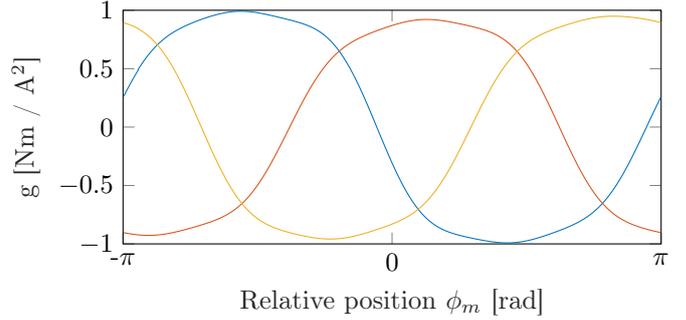}
\vspace*{-7mm}\caption{Example function $g(\phi_m)$. The torque-current relationship is clearly periodic in the rotor angle, and it slightly differs for each of the three coils.}
\label{fig:g}
\end{figure}
First, the nonlinear dynamics of the switched reluctance motor are defined. The 
multi-input single-output (MISO) system is given by
\begin{equation}\label{eq:system}
 P : \left\{ \begin{aligned}
 \dot{x}(t) &= A x(t) + B T(t) \\
 T(t) &= g\left(\phi(t), u(t)\right)\\
\phi(k) &= C x(t),
\end{aligned}\right.
\end{equation}
with $A\in\mathbb{R}^{n\times n}$, $B\in \mathbb{R}^{n\times 1}$, $C\in \mathbb{R}^{1\times n}$. The output $\phi(t)\in \mathbb{R}$ denotes position in radians and the input $u(t)\in \mathbb{R}^{3} \geq 0$ is a vector of squared currents, i.e., $u:=[i_1^2, i_2^2, i_3^2]^\top$. The function $g: \mathbb{R} \times\mathbb{R}^{3} \mapsto \mathbb{R}$ is defined as \begin{equation}
g(\phi, u) := \begin{bmatrix}
g_1(\phi) & g_2(\phi) & g_3(\phi)
\end{bmatrix} u,
\end{equation}
where each $g_c(\phi) : \mathbb{R}\mapsto\mathbb{R}$ is a periodic function with spatial period $p = \frac{2\pi}{n_{\textnormal{teeth}}}$, see Figure \ref{fig:g} for an example. For ease of notation, $g(\phi,u) \equiv g(\phi) u$. Moreover, $g$ is assumed to be known. 

At intervals of $T_s$ seconds, the rotor position is sampled by the ideal sampler $\mathcal{S}$, defined as \begin{equation}
\mathcal{S} : \phi(k) = \phi(k T_s),\quad k\in \{0, 1, \ldots\}.
\end{equation}
Furthermore, the ideal zero-order-hold $\mathcal{H}$ is used to interpolate the input $u(k)$ between samples, i.e., \begin{equation}\label{eq:zoh}
\mathcal{H} : u(k T_s + \delta) = u(k T_s),
\end{equation}
for all $\delta\in[0, T_s)$. 

\begin{figure}[tb]
\centering
\includegraphics[width=\linewidth]{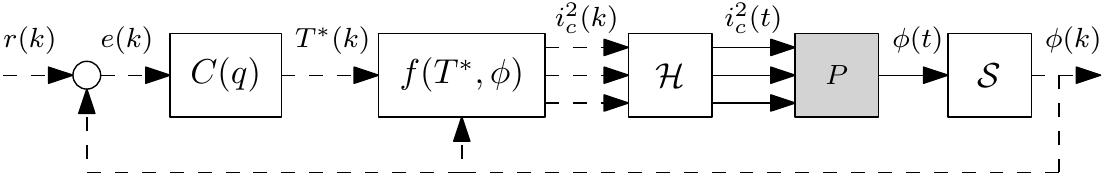}
\vspace*{-3mm}\caption{Closed-loop control scheme of the SRM. Solid lines and dashed lines denote continuous-time and discrete-time signals respectively.}
\label{fig:closed-loop}
\end{figure}

The aim is to construct a commutation function $f:\mathbb{R}\times \mathbb{R}\mapsto\mathbb{R}_{\geq 0}^3$ that linearizes $g(\phi)$, i.e., \begin{equation} \label{eq:invert_simple}
T = g(\phi) f(\phi,T^*) = T^*,
\end{equation}
see Figure \ref{fig:controlscheme}. If \eqref{eq:invert_simple} holds, $T^*$ can be considered to be the input to the linear system $G$ described by state-space matrices $(A,B,C)$, and linear feedback control can be used, see Figure \ref{fig:closed-loop}.

The commutation function $f$ is structured as \begin{equation}\label{eq:fstructure}
f(\phi, T^*) := \begin{bmatrix}
f_1(\phi) & f_2(\phi) & f_3(\phi)
\end{bmatrix}^\top T^*,
\end{equation} 
where each $f_c : \mathbb{R}\mapsto \mathbb{R}_{\geq 0}$ is a periodic function with spatial period $p = \frac{2\pi}{n_{\textnormal{teeth}}}$ that yields a squared current $i_c^2$ for coil $c\in[1,2,3]$ given some requested torque $T^*\in\mathbb{R}$. Note that infinitely many functions $f$ satisfy \eqref{eq:invert_simple} because it is an inner product, and hence, there is design freedom in $f$.
In the next section, it is shown that even if $g$ is known, it is not possible to find a commutation function $f$ that leads to $T(t)=T^*(t)\ \forall t$ in \eqref{eq:invert_simple}, because of sampling effects.
\subsection{Torque ripple due to sampled data aspects}\label{sec:torqueripple}
In practice, torque ripple will always occur on a switched reluctance motor, because the commutation function can only linearize the nonlinear  function $g$ on the samples, but not in between samples. This is explained as follows.

The function $g$ would be perfectly linearized if \eqref{eq:invert_simple} holds at all times, i.e.,
\begin{equation}
T(t) = g(\phi(t)) f(\phi(t)) T^*(t) = T^*(t) ,\quad \forall t.
\end{equation}
This equation cannot be satisfied in practice, because the zero-order-hold scheme in \eqref{eq:zoh} leads to\begin{equation}\begin{aligned}
T(k T_s+\delta) &= g(\phi(k T_s+\delta)) f(\phi(k T_s))T^*(k T_s),
\end{aligned}
\end{equation}
with $\delta \in(0, T_s)$. The produced torque $T(k T_s+\delta)$ does not equal $T^*(k T_s+\delta)$ in general, since \begin{equation}
g(\phi(k T_s+\delta)) f(\phi(k T_s))\neq 1,
\end{equation} 
and therefore, a torque ripple occurs between samples, i.e, $T(t)\neq T^*(t)$ for $t\neq k T_s$. The absolute torque ripple is defined as \begin{equation}\label{eq:absolute_torqueripple}
e_T(t) = T(t)-T^*(t),
\end{equation}
and the relative torque ripple as \begin{equation}\label{eq:relative_torqueripple}\begin{aligned}
e_{T,\text{rel}}(t) := \frac{T(t)-T^*(t)}{T^*(t)}.
\end{aligned}
\end{equation}
In between samples $k$ and $k+1$, i.e., at times $\tau\in [k T_s, (k+1) T_s)$, the relative torque ripple can be written as \begin{equation}\begin{aligned}\label{eq:relative_ripple}
e^{(k)}_{T,\text{rel}}(\tau) &:=\frac{T(\tau)-T^*(k T_s)}{T^*(k T_s)},\\
&= g(\phi(\tau)) f(\phi(k T_s)) - 1, 
\end{aligned}
\end{equation}
so the relative torque error for any given time $t$ is given by \begin{equation}\begin{aligned}\label{eq:relative_ripple_full}
e_{T,\text{rel}}(t) &= e^{(\left\lfloor\frac{t}{T_s}\right\rfloor)}_{T,\text{rel}}(\text{mod}(t,T_s)),\\
&= g(\phi(t)) f\left(\phi\left(\left\lfloor\frac{t}{T_s}\right\rfloor T_s\right)\right) - 1,
\end{aligned}
\end{equation}
where $\lfloor \cdot \rfloor$ denotes the floor operator.

The convolution of the absolute torque ripple with the continuous-time system $G$ will result in a ripple in $\phi(t)$, which may be nonzero on the samples. Consequently, in closed-loop, a periodic position error will occur because of sampling-induced torque ripple. 
\subsection{Problem formulation}\label{sec:requirements}
The above analysis motivates the necessity of taking sampling effects explicitly into account when constructing commutation functions. As such, the aim of this paper is to construct a commutation function $f$, with the structure defined in \eqref{eq:fstructure}, that satisfies the following requirements: \begin{enumerate}[label={R\arabic*:}]
\item The nonlinear function $g$ must be linearized on the samples, i.e., \begin{equation}\label{eq:R1}
g(\phi(k T_s)) f (\phi(k T_s)) = 1.
\end{equation}
\item The 2-norm of the relative torque ripple, i.e., $\|e_{T,\text{rel}}\|_2$ is as small as possible, see \eqref{eq:relative_ripple_full}. Note that $\|e_{T,\text{rel}}\|_2$ will always be greater than zero for $T_s > 0$. 
\item The power consumption of the commutation strategy during constant velocity, given by \begin{equation}
E[f] = \sum_c \|f_c(\phi)\|_1,
\end{equation}
is as small as possible. 
\end{enumerate}
In the next section, it is explained how this problem is solved.

\section{Optimal commutation}\label{sec:optimal}
In this section, it is explained how a commutation function can be constructed that minimizes both sampling-induced torque ripple and power consumption. The key idea is to pose a finite-dimensional optimization problem that explicitly takes R1, R2 and R3 into account, to obtain realizations of an optimal commutation function at a number of rotor angles. Since a commutation function must be continuous, the finite-dimensional solution is used to pose a regression problem, in order to fit a function with some specific structure. In particular, Gaussian Process regression is used. 

First, the optimization problem that addresses the problem posed in Section \ref{sec:requirements} is defined. Subsequently, it is explained how Gaussian Process regression yields a continuous commutation function with a specific model structure. Finally, it is explained how the hyper-parameters of this model can be found automatically. 

\subsection{Optimization problem}
The requirements R1, R2, and R3 posed in Section \ref{sec:requirements} are used to form an optimization problem as follows. The design variables are defined as 
\begin{equation}\label{eq:F}
F := [f({\phi}_{m,1}),...,f(\phi_{m,N})]^\top, 
\end{equation}
where the relative motor angles $\phi_m$ are defined as \begin{equation}
\phi_m:= \phi-2\pi\left\lfloor \frac{\phi+\pi}{2\pi}\right\rfloor,
\end{equation} and $\phi_{m,i}$ are chosen to be regularly spaced between $-\pi$ and $\pi$. Moreover, a nominal velocity is defined as \begin{equation}
v = \frac{2\pi}{n_{\text{t}} N T_s},
\end{equation}
such that when moving at $v$ rad/s, every time $t=k T_s$ corresponds with a relative position $\phi_m(t)=\phi_m(k T_s)=\phi_{m,k}$. At this velocity, samples in time are aligned with the design variables. 

It then follows that R1, see \eqref{eq:R1}, can be written as $N$ equality constraints that are linear in the design variables, namely,
\begin{equation}\label{eq:equality}
g(\phi_{m,i}) f(\phi_{m,i}) = 1,\quad\forall i\in[1,\ldots,N]. 
\end{equation}
Moreover, it must hold that \begin{equation}\label{eq:inequality}
f(\phi_{m,i}) \geq 0,\quad \forall i\in[1,\ldots,N].
\end{equation}
Next, the cost function of the optimization problem is defined, which includes both R2 and R3. The 2-norm of the relative torque ripple \eqref{eq:relative_ripple_full} is approximated numerically by dividing each sample into $M$ subsamples to construct 
\begin{equation}
    e_{T,\text{rel}} = [e_{T,\text{rel}}(0),
     \ldots,
     e_{T,\text{rel}}\left({N T_s - \frac{T_s}{M}}\right),
     e_{T,\text{rel}}({N T_s})
    ]^\top,
    \end{equation}
which is linear in the design variables. 
Finally, R3 is satisfied by including $\|F\|_1$ in the cost function, which is defined as \begin{equation}\begin{aligned}
\|F\|_1 &:= \sum_c \| [f_c({\phi}_{m,1}),...,f_c(\phi_{m,N})]^\top \|_1. \end{aligned}
\end{equation}
This leads to the following convex optimization problem. 
\begin{equation}\label{eq:problem}
\begin{array}{ll}
\min_F & \|F\|_{1}+\beta\left\|e_{T,\text{rel}}\right\|_{2}, \\
\text { subject to } & \eqref{eq:F},\ \eqref{eq:equality},\ \eqref{eq:inequality},
\end{array}
\end{equation}
where $\beta\in\mathbb{R}_{\geq 0}$ is a constant that acts as a trade-off between low power consumption and low torque ripple.

\begin{remark}
Other terms that enforce desired properties on $f$ can be easily added to the cost function, e.g., $\|F\|_{\mathcal{M}}$, where $\mathcal{M}$ defines a norm that enforces smoothness of $f$, or $\|F\|_{\infty}$, such that peak currents are minimized. 
\end{remark}
It is shown next how a continuous commutation function $f$ is created from a finite number of optimal values $F^\star$. 

\subsection{Obtaining a commutation function with Gaussian Process regression}
A commutation function must be continuous, i.e., it must yield squared currents for any rotor angle and requested torque. Since the solution to optimization problem \eqref{eq:problem} contains only $N$ realizations of the optimal commutation function, interpolation is required. The underlying structure of the optimal commutation function is unknown, other than it being periodic in $\phi$. Therefore, a structure is chosen that can represent a large range of functions, based on the smoothness and periodicity of $f$, as follows.

The commutation functions $f_c$ of the coils $c\in[1,2,3]$ are assumed to be of the form \begin{equation}\begin{aligned}\label{eq:structure}
f_c(\phi_m^*) &= \sum_{i=1}^N \alpha_i k_c(\phi_m^*, \phi_{m,i}) = \mathbf{k}_c({\phi_m^*},\Phi_m) \boldsymbol{\alpha}_c,
\end{aligned} 
\end{equation}
where $\phi_m^*$ denotes an arbitrary relative position, $\phi_{m,i}$ are as defined in the previous section, $\boldsymbol{\alpha}\in\mathbb{R}^N$ is a vector of weights and \begin{equation}
\mathbf{k}_c({\phi_m^*},\Phi_m):=[k_c(\phi_m^*,\phi_{m,1}),\allowbreak\ldots,k_c(\phi_m^*,\phi_{m,N})].
\end{equation} The kernel function $k_c(\phi_{m,i},\allowbreak\phi_{m,j})$ imposes some nonlinear structure on $f$, as explained in detail in Section \ref{sec:kernel}. 

The regression problem is then posed as follows. Given a kernel function $k_c$, the weights $\boldsymbol{\alpha}_c$ that best fit the observations $F^\star$ of $f^\star(\phi_m^*)$ are obtained by solving the following GP regression problem for each $c\in[1,2,3]$:
\begin{equation}\begin{aligned}
\min _{\boldsymbol{\alpha}_c} J[f_c]=\|F_c^\star-K_c \boldsymbol{\alpha}_c\|_{2}^{2}+\sigma_{n,c}^{2} \boldsymbol{\alpha}_c^{\top} K_c \boldsymbol{\alpha}_c,
\end{aligned}
\end{equation}
where $K_c$ is shorthand for the Gramian $K_c(\Phi_m,\Phi_m)$ and $\sigma_{n,c}^2>0$ is the regularization strength. This problem has the solution \begin{equation}
\hat{\boldsymbol{\alpha}}_c = (K_c+\sigma_{n,c}^2 I) F_c^\star,
\end{equation}
and hence, the continuous commutation function $f_c$, which is defined for an arbitrary $\phi_m^*$ is given by \begin{equation}\label{eq:posterior}
f_c(\phi_m^*) = \mathbf{k}_c({\phi_m^*},\Phi_m) \hat{\boldsymbol{\alpha}}_c.
\end{equation}
The next section explains the choice of kernel function $k_c$. 
\subsection{Kernel function}\label{sec:kernel}
From \eqref{eq:structure}, it is clear that the kernel function $k_c$ imposes structure, or a prior, on the function space of $f_c$. A convenient choice that allows for a large degree of flexibillity is to pose a prior on the degree of smoothness of $f_c$, this is done with a Matèrn kernel \citep{Rasmussen2004} of the form \begin{equation}\begin{aligned}\label{eq:kernel}
k_{c,\mu/2}(x,x^\prime)&=\sigma_f^{2} \exp \left(-{\sqrt{2 \mu+1} \rho}\right) \frac{\mu !}{(2 \mu) !} \\ &\cdot\sum_{i=0}^{\mu} \frac{(\mu+i) !}{i !(\mu-i) !}\left({2 \sqrt{2 \mu+1} \rho}\right)^{\mu-i},
\end{aligned}
\end{equation}
where $\mu\in\mathbb{N}$, and $\rho = \sqrt{{(x^\top x^\prime)}/{\ell^2}}$. A function $f_c(\phi_m^*)$ parametrized by this kernel through \eqref{eq:posterior} is $\mu$-times differentiable in the mean-square sense, and hence, the parameter $\mu$ relates to the smoothness of $f_c$. 
Moreover, it is known that the commutation function is periodic in $\phi$ with spatial period $p=\frac{2\pi}{n_{\text{t}}}$, and this periodicity is enforced on $f_c$ by $k_c$ by defining \begin{equation}
x = \begin{bmatrix}
\sin\frac{2\pi\phi_{m}}{p}\\ \cos\frac{2\pi\phi_{m}}{p}
\end{bmatrix},\quad x^\prime = \begin{bmatrix}
\sin\frac{2\pi\phi_{m}^\prime}{p}\\ \cos\frac{2\pi\phi_{m}^\prime}{p}
\end{bmatrix},
\end{equation}
see \cite[Section 2.7]{Duvenaud2014} for further details. The parameters $\mu$, $\ell$, $\sigma_f^2$ and $\sigma_{n,c}^2$ can be tuned automatically, as is explained next.
\subsection{Optimization of hyper-parameters}
The model structure of the commutation function in \eqref{eq:posterior} is a function of the kernel, and hence, it depends on the kernel hyper-parameters $\Theta := \{\mu, \ell, \sigma_f^2, \sigma_{n,c}^2\}$. To find a model structure of $f_c$ that is most consistent with the realizations $F_c^\star$ following from \eqref{eq:problem}, the marginal likelihood is maximized. The marginal likelihood $p(F_c^\star \mid \Phi_m, \Theta)$ is the probability of the data given the model, parametrized by \eqref{eq:kernel} in terms of $\Theta$. Its logarithm is given by
\begin{equation}\begin{aligned}\label{eq:marginal}
\text{log}\ p(F_c^\star \mid \Phi_m, \Theta) &= -\frac{1}{2} (F_c^\star)^\top (\tilde{K}_{c}(\Theta))^{-1} F_c^\star \\&- \frac{1}{2} \log \det (\tilde{K}_c(\Theta))-\frac{N}{2}\log 2\pi,\end{aligned}
\end{equation}
with $\tilde{K}_c(\Theta) = K_c+\sigma_{n,c}^2 I$. By maximizing this expression with respect to $\Theta$, the hyper-parameters $\Theta^\star$ are found such that commutation function $f_c(\phi_m^*, \Theta^\star)$ in \eqref{eq:posterior} is the most likely fit through $F^\star$ given the kernel structure. 
\section{Simulation results}\label{sec:sim}
The method is applied to a simulation example to demonstrate its effectiveness in reducing torque ripple. The setting is explained first, after which the results are presented.
\subsection{Setting}
Consider the control scheme in Figure \ref{fig:controlscheme}. The system, consisting of a mass and a damper, is described by \begin{equation}\label{eq:plant}
G(s) = \frac{1}{s(s+1)},
\end{equation}
and the controller, with sampling rate $F_s=1000$ Hz, by \begin{equation}
C(z) = \frac{6.72\cdot 10^5 z^2 - 1.1\cdot 10^6 z + 4.51\cdot 10^5}{z^2 - 1.03 z + 0.0296},
\end{equation}
such that the bandwidth is 100 Hz. The function $g$ is shown in Figure \ref{fig:g}, where the torque-current relationship is deliberately defined differently for every coil, see also \citet{Kramer2020}. The rotor of the SRM consists of $n_{\text{t}}=131$ teeth. The goal is to track a number of second-order references which consist of a constant acceleration part during 5 teeth, and a constant velocity part during 15 teeth, where $v_j$ is different for every reference. 
\subsection{Approach}
The optimization problem \eqref{eq:problem} is posed at $N=150$ points with $M=15$ subsamples per sample. The parameter $\beta$ that trades off power consumption and sampling-induced torque ripple is chosen as $\beta=1000$, and \eqref{eq:problem} is solved with SeDuMi, see \cite{Sturm1999}. Subsequently, for every coil a kernel in the form of \eqref{eq:kernel} is defined with $\mu=3$ and $p=\frac{2\pi}{n_{
\text{t}}}$. The parameters $\ell,\ \sigma_f^2$ and $\sigma_{n,c}^2$ are found by maximization of \eqref{eq:marginal}. With these kernels, a commutation function is defined through \eqref{eq:posterior}. The computational time required for solving for $F^\star$, optimizing the hyper-parameters and computing $\hat{\boldsymbol{\alpha}}$ is less than 10 s on a personal computer. 
\subsection{Results}
\begin{figure}[tb]
\centering
\input{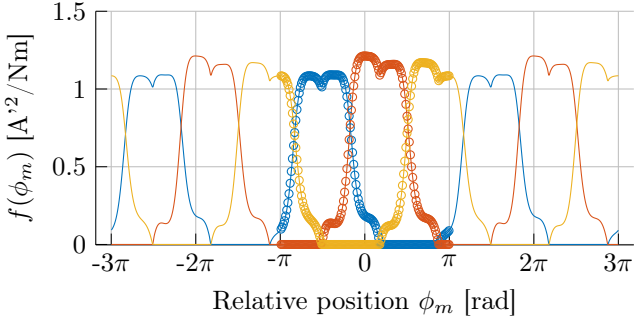}
\vspace*{-3mm}\caption{Solution $F^\star$ of optimization problem \eqref{eq:problem} (\protect\bluedot, \protect\reddot, \protect\yeldot), with the continuous Gaussian Process fit $f^\star(\phi_m^*)$ (\protect\blueline, \protect\redline, \protect\yelline). Clearly, $f$ is periodic in $\phi_m$ and it is fitted well through $F^\star$. The commutation function is different for every coil because $g_c$ is different for every coil.}
\label{fig:gp_fig}
\end{figure}
\begin{figure}[tb]
\centering
\input{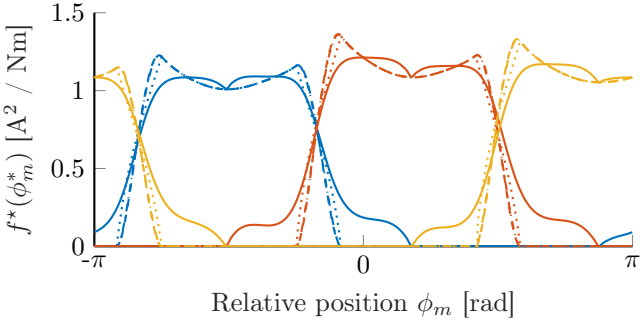}
\vspace*{-3mm}\caption{Comparison of the developed commutation function $f^\star(\phi_m)$ using $\beta=1000$ (solid) with the conventional functions $f_{\text{sine}}$ (dashed), $f_{\text{cubic}}$ (dot-dashed) and $f_{\text{lin}}$ (dotted).}
\label{fig:compare_conventional}
\end{figure}
The resulting commutation function is shown in Figure \ref{fig:gp_fig}, and 
Figure \ref{fig:compare_conventional} compares the resulting commutation function with conventional functions. In particular, it appears to be fundamentally different from the squared-sinusoidal commutation function $f_{\text{sine}}$ \citep{Vujicic2012,Kramer2020}, as well as functions $f_{\text{cubic}}$ and $f_{\text{lin}}$, see \citet{Wang2016}. Each of these functions are of the form  
\begin{equation}\label{eq:tsf}
    f_{\text{conv},c}(\phi_m) = 
    f_{\text{TSF}}\left(\phi_m + \frac{2\pi(c-1)}{3}\right) \text{sat}(1/g_c(\phi_m)) 
    \end{equation}
where $f_{\text{TSF}}(\phi_m)$ is a so-called torque sharing function \citep{Ilic-Spong1987} which is periodic in $\phi_m$, and sat$(x)$ saturates $x$ between $[-a,a],\ a\in\mathbb{R}$. Note that for angles where saturation is required, which is the case for any $\phi_m$ such that $g_c(\phi_m)\approx 0$, it does not hold that $g(\phi_m) f(\phi_m)=1$, i.e., R1 is not satisfied in general by commutation functions based on torque sharing. To prevent the need for saturation as much as possible, the overlap is typically chosen to be small, since then $f(\phi_m)=0$ at positions where $g(\phi_m)\approx 0$. In this example, the overlap between the commutation functions is chosen to be $\phi_{m,ov}=\frac{\pi}{6}$ radians, and the saturation level is chosen to be $a=3$ A$^2$/Nm.

In contrast to conventional commutation functions, the optimal commutation function is not constructed through piece-wise division by $g_c(\phi_m),\ c\in[1,2,3]$ as in \eqref{eq:tsf}, and hence, a large degree of overlap can be achieved while still satisfying R1. 
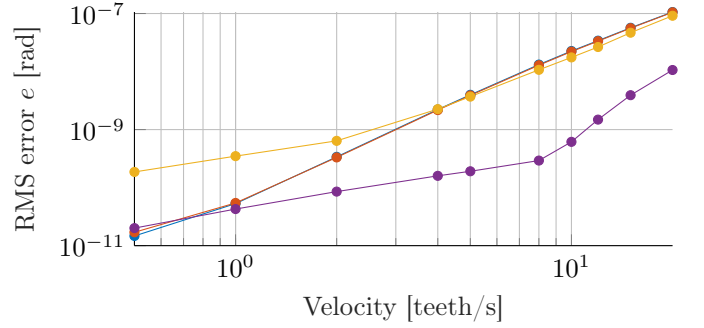
\begin{figure}[tb]
\input{different_speeds.tex}
\vspace*{-7mm}\caption{Closed-loop position error using conventional commutation functions $f_{\text{sine}}$ (\protect\blueline), $f_{\text{cubic}}$ (\protect\redline), $f_{\text{lin}}$ (\protect\yelline), and the optimal commutation function $f^\star$ (\protect\purline).}
\label{fig:different_speeds}
\end{figure} 

Next, the closed-loop performance of $f^\star$ is compared to conventional approaches, see Figure \ref{fig:different_speeds}. The position error is reduced significantly for a range of velocities. At low velocities, the effect of torque ripple is negligible, since then $\phi_m(\tau)\approx\phi_m(k T_s)$ for a given $T_s$, and hence $e_{T,\text{rel}}^{(k)}(\tau)$ is small in \eqref{eq:relative_ripple}. This is confirmed by the results in Figure \ref{fig:different_speeds}. Consequently, the performance increase is most pronounced at higher velocities. 
\begin{figure}[tb]
\centering
\input{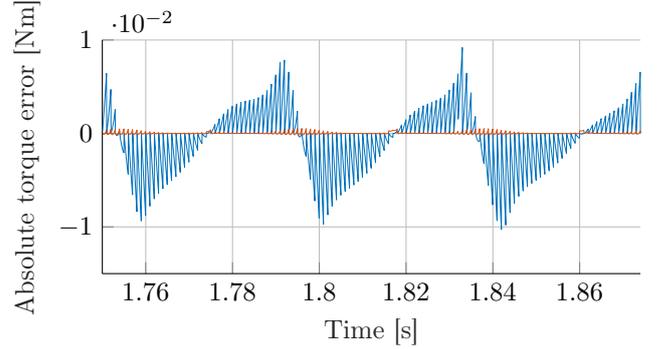}
\vspace*{-3mm}\caption{Absolute torque error $T^*(t)-T(t)$ during the last tooth of a closed-loop simulation at $v=8$ teeth/s, using conventional commutation function $f_{\text{sine}}$ (\protect\blueline) and $f^\star$ (\protect\redline), that was optimized specifically to minimize torque ripple. The torque error at $t=k T_s$ is always zero, but the optimal commutation function reduces the inter-sample torque ripple.}
\label{fig:Result_Terr}
\end{figure}
\begin{figure}[tb]
\centering
\input{Result_err.tex}
\vspace*{-3mm}\caption{Position error during the last tooth of a closed-loop simulation at $v=8$ teeth/s, using a squared sinusoidal commutation function (\protect\blueline) and the optimal commutation function (\protect\redline).} 
\label{fig:Result_err}
\end{figure}
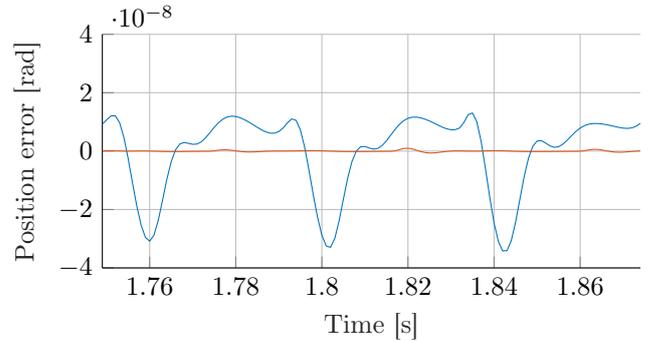

The torque errors resulting from $f^\star$ and $f_{\text{sine}}$ are compared in Figure \ref{fig:Result_Terr} for $v=8$ teeth/s. Note that the horizontal axis is continuous-time, such that it includes inter-sample torque ripple. The optimal commutation function leads to significantly less torque ripple between samples, and as a result, the position error is reduced considerably, see Figure \ref{fig:Result_err}. Clearly, the reduction in inter-sample torque ripple leads to an increase in tracking performance.
\subsection{Torque ripple vs. energy consumption}
\begin{figure}[tb]
\centering
\input{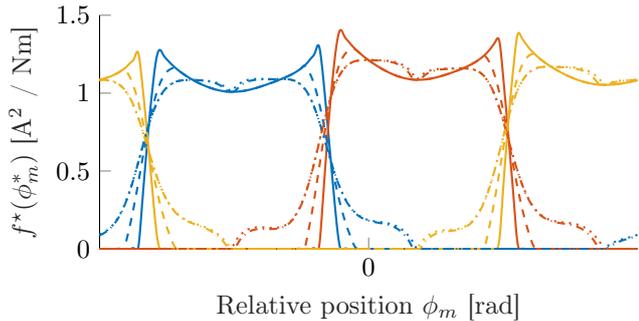}
\vspace*{-3mm}\caption{Optimal commutation functions $f^\star(\phi_m)$ resulting from $\beta=1$ (solid), $\beta=1.5$ (dashed), $\beta=2.2$ (dash-dotted) and $\beta=1000$ (dotted). Higher values of $\beta$ correspond with a higher emphasis on preventing torque ripple in relation to power consumption. The function is different for every coil because $g_c$ is different for every coil.}
\label{fig:compare_optimal} 
\end{figure}

A tuning knob in the developed approach is $\beta$, which trades off power consumption and torque ripple in \eqref{eq:problem}. Figure \ref{fig:compare_optimal} compares some commutation functions for different values of $\beta$. As expected, it can be seen that a larger emphasis on power consumption, corresponding to a small $\beta$, yields commutation functions with a smaller area.

\begin{figure}[tb]
\centering
\input{beta_combined.tex}
\vspace*{-7mm}\caption{Performance increase (\protect\blueline, left axis) and energy increase (\protect\redline, right axis) of the developed commutation function at $v=8$ teeth/s with respect to a squared-sinusoidal commutation function. Higher values of $\beta$ correspond with a larger emphasis on reducing sampling-induced torque ripple.}
\label{fig:beta_combined}
\end{figure}
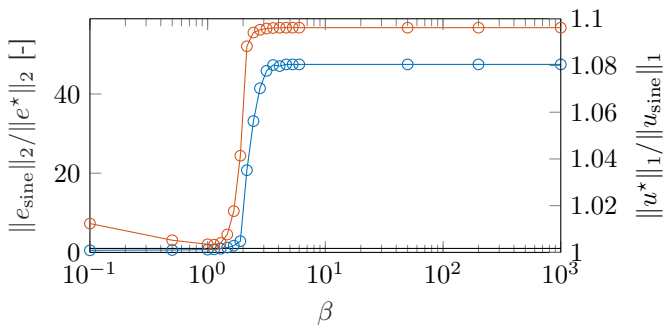

Moreover, Figure \ref{fig:beta_combined} shows the performance increase with respect to $f_{\text{sine}}$, for different values of $\beta$ at $v=8$ teeth/s. As expected from \eqref{eq:problem}, the torque ripple is reduced most for large values of $\beta$ and consequently, the reduction in the position error is most pronounced, at the cost of increased energy usage. Note that the energy consumption is evaluated in closed-loop, and thus, if low values of $\beta$ lead to large errors, the feedback controller will lead to more energy consumption as well. 

From these results it is clear that by optimizing a commutation function explicitly to reduce inter-sample torque ripple, a function is obtained that is fundamentally different from conventional commutation functions, leading to significantly better performance in closed-loop. The parameter $\beta$, that trades of power consumption and torque ripple, is an intuitive tuning knob of the approach. This parameter is shown to influence the degree of overlap in currents between different coils, which is in contrast with conventional approaches, where the degree of overlap is chosen in an ad-hoc manner. The results offer a new perspective on solutions to the over-parametrized commutation problem, and suggest that the developed approach is arguably more methodical and intuitive than conventional approaches based on torque sharing functions. 

\section{Conclusion and future work}\label{sec:conclusion}
In this paper, it is shown that sampling effects are a cause of torque ripple in switched reluctance motors, and that this ripple can be significantly reduced by exploiting the design freedom inherent in the commutation problem. The resulting commutation functions are fundamentally different from conventional functions found in literature, leading to a factor 45 decrease in the 2-norm of the error in simulation. The performance increase is especially pronounced at high velocities or low sampling rates. 

As in most commutation strategies, the design of the commutation function relies on a model of the nonlinear relationship $g$ between torque, current and angle. Since the developed approach exploits knowledge of $g$ in the optimization problem, the effectiveness in practice depends on the model quality. Therefore, future work will be aimed towards data-driven design of commutation functions. 

\bibliography{library.bib}

\end{document}

%% file: tikz_set.tex
\usepackage{tikz}
\usepackage{tikzscale}

\definecolor{myorange}{cmyk}{0,0.35,0.85,0} 
\definecolor{mypurple}{cmyk}{0.5,1,0,0} 

\definecolor{matblue1}{rgb}{0,0.4470,0.7410}
\definecolor{matred1}{rgb}{0.85,0.325,0.098}
\definecolor{matyel1}{rgb}{0.9290, 0.6940, 0.1250}
\definecolor{matpur1}{rgb}{0.4940, 0.1840, 0.5560}
\definecolor{matgre1}{rgb}{0.4660, 0.6740, 0.1880}
\definecolor{matblue2}{rgb}{0.3010, 0.7450, 0.9330}
\definecolor{matred2}{rgb}{0.6350, 0.0780, 0.1840}
\definecolor{matgrey1}{rgb}{0.5, 0.6, 0.7}
\definecolor{matpink1}{rgb}{1, 0.07, 0.65}
\definecolor{matblue3}{rgb}{0.07, 0.62, 1}

    \definecolor{mblue}{rgb}{0,0.447,0.741}
    \definecolor{mred}{rgb}{0.85,0.325,0.098}
    \definecolor{myellow}{rgb}{0.9290,0.6940,0.1250}
    \definecolor{mmagenta}{rgb}{1,0,1}
    \definecolor{mgreen}{rgb}{0.4460,0.6740,0.1880}
    \definecolor{mgrey}{rgb}{0.6,0.6,0.6}
    \definecolor{mpurple}{rgb}{0.4940, 0.1840, 0.5560}
    
    \tikzset{cross/.style={cross out, draw=black, minimum size=2*(#1-\pgflinewidth), inner sep=0pt, outer sep=0pt}, cross/.default={1pt}}
    
    \usetikzlibrary{shapes}

\newcommand{\blueline}{\raisebox{2pt}{\tikz{\draw[-,matblue1,solid,line width = 0.9pt](0,0) -- (3mm,0);}}}
\newcommand{\redline}{\raisebox{2pt}{\tikz{\draw[-,matred1,solid,line width = 0.9pt](0,0) -- (3mm,0);}}}

\newcommand{\bluedot}{\raisebox{.7pt}{\tikz{\draw[-,matblue1,solid,line width = 1pt](0,0) circle (.7mm);}}}
\newcommand{\reddot}{\raisebox{.7pt}{\tikz{\draw[-,matred1,solid,line width = 1pt](0,0) circle (.7mm);}}}
\newcommand{\yeldot}{\raisebox{.7pt}{\tikz{\draw[-,matyel1,solid,line width = 1pt](0,0) circle (.7mm);}}}

\newcommand{\yelline}{\raisebox{2pt}{\tikz{\draw[-,matyel1,solid,line width = 0.9pt](0,0) -- (3mm,0);}}}
\newcommand{\purline}{\raisebox{2pt}{\tikz{\draw[-,matpur1,solid,line width = 0.9pt](0,0) -- (3mm,0);}}}

%% file: different_speeds.tex
%
%
\definecolor{mycolor1}{rgb}{0.00000,0.44700,0.74100}%
\definecolor{mycolor2}{rgb}{0.85000,0.32500,0.09800}%
\definecolor{mycolor3}{rgb}{0.92900,0.69400,0.12500}%
\definecolor{mycolor4}{rgb}{0.49400,0.18400,0.55600}%
\begin{tikzpicture}

\begin{axis}[%
width=0.8\linewidth,
height=0.35\linewidth,
at={(0in,0in)},
scale only axis,
xmode=log,
xmin=0.5,
xmax=20,
xminorticks=true,
xlabel style={font=\color{white!15!black}},
xlabel={Velocity [teeth/s]},
ymode=log,
ymin=1e-11,
ymax=1.06132044872361e-07,
yminorticks=true,
ytick={1e-11, 1e-9, 1e-7},
ylabel style={font=\color{white!15!black}},
ylabel={RMS error $e$ [rad]},
axis background/.style={fill=white},
axis x line*=bottom,
axis y line*=left,
xmajorgrids,
xminorgrids,
ymajorgrids,
yminorgrids
]
\addplot [color=mycolor1, mark size=1.7pt, mark=*, mark options={solid, mycolor1}, forget plot]
  table[row sep=crcr]{%
0.5	1.46858567992884e-11\\
1	5.2987128277835e-11\\
2	3.3909860439395e-10\\
4	2.22078562854584e-09\\
5	3.97657362441298e-09\\
8	1.31908491812556e-08\\
10	2.26033568288798e-08\\
12	3.40802156664718e-08\\
15	5.67666506165803e-08\\
20	1.06132044872361e-07\\
};
\addplot [color=mycolor2, mark size=1.7pt, mark=*, mark options={solid, mycolor2}, forget plot]
  table[row sep=crcr]{%
0.5	1.69982923046691e-11\\
1	5.4434424930479e-11\\
2	3.31312114674291e-10\\
4	2.16845485537898e-09\\
5	3.88006347494586e-09\\
8	1.28576516616307e-08\\
10	2.20822638306579e-08\\
12	3.34094657763568e-08\\
15	5.58611843069892e-08\\
20	1.05000323366135e-07\\
};
\addplot [color=mycolor3, mark size=1.7pt, mark=*, mark options={solid, mycolor3}, forget plot]
  table[row sep=crcr]{%
0.5	1.86318263111989e-10\\
1	3.48635611618119e-10\\
2	6.39783185933449e-10\\
4	2.2524721598822e-09\\
5	3.69802688748789e-09\\
8	1.0715067533921e-08\\
10	1.75375925310113e-08\\
12	2.64895441952849e-08\\
15	4.67370272486634e-08\\
20	9.15688885211661e-08\\
};
\addplot [color=mycolor4, mark size=1.7pt, mark=*, mark options={solid, mycolor4}, forget plot]
  table[row sep=crcr]{%
0.5	2.00546007221358e-11\\
1	4.27119741827701e-11\\
2	8.53663843007374e-11\\
4	1.59598411442411e-10\\
5	1.91253315267128e-10\\
8	2.92552925725944e-10\\
10	6.14893405704665e-10\\
12	1.48539233432142e-09\\
15	3.91591823498744e-09\\
20	1.06355871271576e-08\\
};
\end{axis}
\end{tikzpicture}%

%% file: Result_err.tex
%
%
\definecolor{mycolor1}{rgb}{0.00000,0.44700,0.74100}%
\definecolor{mycolor2}{rgb}{0.85000,0.32500,0.09800}%
\begin{tikzpicture}

\begin{axis}[%
width=0.8\linewidth,
height=0.35\linewidth,
at={(0.768in,0.425in)},
scale only axis,
xmin=1.749,
xmax=1.874,
xlabel style={font=\color{white!15!black}},
xlabel={Time [s]},
ymin=-4e-08,
ymax=4e-08,
ylabel style={font=\color{white!15!black}},
ylabel={Position error [rad]},
axis background/.style={fill=white},
axis x line*=bottom,
axis y line*=left,
xmajorgrids,
ymajorgrids
]
\addplot [color=mycolor1, forget plot]
  table[row sep=crcr]{%
1.749	9.29327215182241e-09\\
1.75	1.06976421054839e-08\\
1.751	1.20514843660047e-08\\
1.752	1.21165830702097e-08\\
1.753	9.77595271400844e-09\\
1.754	4.69094241228873e-09\\
1.755	-2.57373633516522e-09\\
1.756	-1.08463739101339e-08\\
1.757	-1.87991459066339e-08\\
1.758	-2.53024761143195e-08\\
1.759	-2.95358836277515e-08\\
1.76	-3.08396227444163e-08\\
1.761	-2.87589574377023e-08\\
1.762	-2.36398213226963e-08\\
1.763	-1.67156093411336e-08\\
1.764	-9.58316492827294e-09\\
1.765	-3.61476071297773e-09\\
1.766	4.21921608761977e-10\\
1.767	2.45437259405179e-09\\
1.768	2.95541180417302e-09\\
1.769	2.66688349181976e-09\\
1.77	2.31733388034172e-09\\
1.771	2.42058506572107e-09\\
1.772	3.19192350328734e-09\\
1.773	4.57236692952279e-09\\
1.774	6.32216157203658e-09\\
1.775	8.13596590099053e-09\\
1.776	9.74044578327948e-09\\
1.777	1.0952393769692e-08\\
1.778	1.16939598004961e-08\\
1.779	1.19752368021864e-08\\
1.78	1.18607300647611e-08\\
1.781	1.14356334401933e-08\\
1.782	1.07826577666614e-08\\
1.783	9.97335480867889e-09\\
1.784	9.07193054011657e-09\\
1.785	8.14596057274741e-09\\
1.786	7.27752302864815e-09\\
1.787	6.5696428297457e-09\\
1.788	6.14556727729365e-09\\
1.789	6.14127670939268e-09\\
1.79	6.69385735729833e-09\\
1.791	7.93080889849307e-09\\
1.792	9.69643232373585e-09\\
1.793	1.10328607361154e-08\\
1.794	1.05421513740822e-08\\
1.795	7.30215332644946e-09\\
1.796	1.32740152203326e-09\\
1.797	-6.52284237734335e-09\\
1.798	-1.49620565936814e-08\\
1.799	-2.27237342276254e-08\\
1.8	-2.8814825525636e-08\\
1.801	-3.24764714010328e-08\\
1.802	-3.29774439977726e-08\\
1.803	-2.98983127144936e-08\\
1.804	-2.3880510346963e-08\\
1.805	-1.64108598932344e-08\\
1.806	-9.18319875697904e-09\\
1.807	-3.50699647100328e-09\\
1.808	1.47082346302341e-11\\
1.809	1.49963108597717e-09\\
1.81	1.57308277515256e-09\\
1.811	1.05750319612241e-09\\
1.812	6.87113810293738e-10\\
1.813	9.26433041392727e-10\\
1.814	1.91669702331865e-09\\
1.815	3.52864559793886e-09\\
1.816	5.47491818547741e-09\\
1.817	7.43185413210057e-09\\
1.818	9.13319053719164e-09\\
1.819	1.04181140381243e-08\\
1.82	1.12350656555904e-08\\
1.821	1.16152462092955e-08\\
1.822	1.16342111500245e-08\\
1.823	1.1377500719334e-08\\
1.824	1.09198247111308e-08\\
1.825	1.03200326062591e-08\\
1.826	9.62836133044931e-09\\
1.827	8.89943396753523e-09\\
1.828	8.20431944603683e-09\\
1.829	7.63685070737097e-09\\
1.83	7.31231430961543e-09\\
1.831	7.35943728180644e-09\\
1.832	7.90860699151352e-09\\
1.833	9.84293868633301e-09\\
1.834	1.24885576324729e-08\\
1.835	1.30579710377887e-08\\
1.836	1.00215429288397e-08\\
1.837	3.29379334917945e-09\\
1.838	-5.94815985355268e-09\\
1.839	-1.58603299382776e-08\\
1.84	-2.46270713821772e-08\\
1.841	-3.09537520060132e-08\\
1.842	-3.41935607606203e-08\\
1.843	-3.4103432411392e-08\\
1.844	-3.06349631218339e-08\\
1.845	-2.43078609463154e-08\\
1.846	-1.64426482429647e-08\\
1.847	-8.67759464284745e-09\\
1.848	-2.39206510155299e-09\\
1.849	1.66833480363948e-09\\
1.85	3.47604123263068e-09\\
1.851	3.56347318319905e-09\\
1.852	2.73790123905826e-09\\
1.853	1.79121828480788e-09\\
1.854	1.29270505411228e-09\\
1.855	1.50208601112922e-09\\
1.856	2.39328046269804e-09\\
1.857	3.74932107582993e-09\\
1.858	5.27988430754078e-09\\
1.859	6.72112265842628e-09\\
1.86	7.89520959720846e-09\\
1.861	8.72595062872961e-09\\
1.862	9.22090370725215e-09\\
1.863	9.43698352795508e-09\\
1.864	9.44592681850764e-09\\
1.865	9.31068278031688e-09\\
1.866	9.07682584649905e-09\\
1.867	8.77701966661704e-09\\
1.868	8.44293746027347e-09\\
1.869	8.11814193646399e-09\\
1.87	7.8667853342651e-09\\
1.871	7.77551267816534e-09\\
1.872	7.94868104669888e-09\\
1.873	8.49904535638046e-09\\
1.874	9.53711276618918e-09\\
};
\addplot [color=mycolor2, forget plot]
  table[row sep=crcr]{%
1.749	9.24339493835191e-11\\
1.75	1.50138013133017e-10\\
1.751	1.38402511673519e-10\\
1.752	9.11118958057955e-11\\
1.753	4.06196187796581e-11\\
1.754	8.06688049692639e-12\\
1.755	5.62994095787417e-13\\
1.756	1.29908306334414e-11\\
1.757	3.2529756666122e-11\\
1.758	4.49760229059848e-11\\
1.759	4.05757649701854e-11\\
1.76	1.70740088734078e-11\\
1.761	-2.08204564700054e-11\\
1.762	-6.43832764879448e-11\\
1.763	-1.04350195151426e-10\\
1.764	-1.33667410473493e-10\\
1.765	-1.48886680761962e-10\\
1.766	-1.50216283856253e-10\\
1.767	-1.40651490454502e-10\\
1.768	-1.24607657525644e-10\\
1.769	-1.06474939975953e-10\\
1.77	-8.94720963984241e-11\\
1.771	-7.49864614846274e-11\\
1.772	-6.30356877806548e-11\\
1.773	-4.80114836776124e-11\\
1.774	-3.16453530047056e-11\\
1.775	9.50907130814471e-11\\
1.776	3.12714520944724e-10\\
1.777	4.39387082273868e-10\\
1.778	4.30058100242547e-10\\
1.779	2.89639423556309e-10\\
1.78	8.34579072517272e-11\\
1.781	-1.14995457600742e-10\\
1.782	-2.52101339803801e-10\\
1.783	-3.06160208296546e-10\\
1.784	-2.85125034693579e-10\\
1.785	-2.15592654839725e-10\\
1.786	-1.29378618929366e-10\\
1.787	-5.25980370369439e-11\\
1.788	2.34035013590983e-13\\
1.789	2.69120281615187e-11\\
1.79	3.44079209568804e-11\\
1.791	3.39879235866647e-11\\
1.792	3.63967744831939e-11\\
1.793	4.8246850958833e-11\\
1.794	7.03307412308618e-11\\
1.795	9.81976722158606e-11\\
1.796	1.24667276502066e-10\\
1.797	1.42341693987191e-10\\
1.798	1.45037537535586e-10\\
1.799	1.29130595105664e-10\\
1.8	9.49256229176854e-11\\
1.801	4.70050665057897e-11\\
1.802	-7.07534031363366e-12\\
1.803	-5.90675286460396e-11\\
1.804	-1.0212364287554e-10\\
1.805	-1.32124422513868e-10\\
1.806	-1.48081102935294e-10\\
1.807	-1.51649914847951e-10\\
1.808	-1.46082701490968e-10\\
1.809	-1.35082167673772e-10\\
1.81	-1.21887167026102e-10\\
1.811	-1.08732578496529e-10\\
1.812	-9.66318136619293e-11\\
1.813	-8.5949469763591e-11\\
1.814	-7.46177564181494e-11\\
1.815	-6.35576036245311e-11\\
1.816	-5.62266899706287e-11\\
1.817	8.98646712599316e-11\\
1.818	5.00039343265257e-10\\
1.819	9.03537134000487e-10\\
1.82	9.85157955213367e-10\\
1.821	7.3858463789378e-10\\
1.822	2.9742741602945e-10\\
1.823	-1.61852531377349e-10\\
1.824	-5.00319785601278e-10\\
1.825	-6.52347620366811e-10\\
1.826	-6.25135165854829e-10\\
1.827	-4.75551820144915e-10\\
1.828	-2.78681189236352e-10\\
1.829	-1.00249253343065e-10\\
1.83	1.99490424179771e-11\\
1.831	7.19531101367465e-11\\
1.832	6.94837520853753e-11\\
1.833	3.8600123097865e-11\\
1.834	6.05449024249083e-12\\
1.835	-9.28079835205153e-12\\
1.836	-2.96318525272454e-13\\
1.837	2.92972313076234e-11\\
1.838	6.89309720414144e-11\\
1.839	1.05429109886757e-10\\
1.84	1.26650356868652e-10\\
1.841	1.24755983321734e-10\\
1.842	9.85547199405801e-11\\
1.843	5.35332889128881e-11\\
1.844	-8.22009127432466e-13\\
1.845	-5.45484768466054e-11\\
1.846	-9.97637528143969e-11\\
1.847	-1.31832433858392e-10\\
1.848	-1.49402601401505e-10\\
1.849	-1.5391810048726e-10\\
1.85	-1.48790424425727e-10\\
1.851	-1.3821455091545e-10\\
1.852	-1.25875643242068e-10\\
1.853	-1.13994147454832e-10\\
1.854	-1.03304587106834e-10\\
1.855	-9.38060740196534e-11\\
1.856	-8.53835890879395e-11\\
1.857	-7.81166242802556e-11\\
1.858	-7.19159176654216e-11\\
1.859	-5.65893998327738e-11\\
1.86	-3.49565931756501e-11\\
1.861	1.2641176994066e-10\\
1.862	4.14156264838539e-10\\
1.863	6.02196070786931e-10\\
1.864	5.92471516291937e-10\\
1.865	3.98651889277346e-10\\
1.866	1.16112230941212e-10\\
1.867	-1.52460599700532e-10\\
1.868	-3.32543104164529e-10\\
1.869	-3.9504033377824e-10\\
1.87	-3.52729623287473e-10\\
1.871	-2.44634978940894e-10\\
1.872	-1.16833653862614e-10\\
1.873	-6.61837251669795e-12\\
1.874	6.6005756416132e-11\\
};
\end{axis}
\end{tikzpicture}%

%% file: beta_combined.tex
%
%
\definecolor{mycolor1}{rgb}{0.00000,0.44700,0.74100}%
\begin{tikzpicture}

\begin{axis}[%
width=0.7\linewidth,
height=0.35\linewidth,
at={(0in,0in)},
axis y line*=left,
scale only axis,
xmode=log,
xmin=0.1,
xmax=1000,
xminorticks=true,
xlabel style={font=\color{white!15!black}},
xlabel={$\beta$},
ymin=0,
ymax=59,
ylabel style={font=\color{white!15!black}},
ylabel={$\|e_{\text{sine}}\|_2 / \|e^\star\|_2$ [-]},
axis background/.style={fill=white}
]
\addplot [color=mycolor1, mark=o, mark options={solid, mycolor1}, forget plot]
  table[row sep=crcr]{%
0.1	0.524483867127062\\
0.5	0.601902733530378\\
1	0.698921732987472\\
1.13653347600972	0.777233192430669\\
1.29170834209075	0.929688974138703\\
1.46806977202715	1.21581229782455\\
1.66851044102683	1.70010786366136\\
1.89631797129874	2.83429881619366\\
2.15522885553986	20.7618806564208\\
2.44948974278318	33.1517278422028\\
2.78392709181553	41.4443051687304\\
3.16402633461875	45.84709326885\\
3.59602184827055	47.3103061397156\\
4.08699921102184	47.0555063625313\\
4.64501141975166	47.4743725757597\\
5.27921097499522	47.4743725757597\\
6	47.4743725757597\\
50	47.4743725757597\\
200	47.4743725757597\\
1000	47.4743725757597\\
};
\addplot [color=black, forget plot]
  table[row sep=crcr]{%
0.1	1\\
1000	1\\
};
\end{axis}

\begin{axis}[%
    axis y line*=right,
  axis x line=none,
xmin=0.1,
xmax=1000,
xmode=log,
ymin=1,
ymax=1.1,
width=0.7\linewidth,
height=0.35\linewidth,
scale only axis,
ylabel={$\|u^\star\|_1 / \|u_{\text{sine}}\|_1$},
]
\addplot [color=mred, mark=o, mark options={solid, mred}, forget plot]
  table[row sep=crcr]{%
0.1	1.01234332284091\\
0.5	1.00521231361457\\
1	1.00347257943809\\
1.13653347600972	1.00331136601775\\
1.29170834209075	1.00404943413402\\
1.46806977202715	1.00765772692774\\
1.66851044102683	1.01765927141913\\
1.89631797129874	1.04137666466232\\
2.15522885553986	1.08822362884607\\
2.44948974278318	1.09410067856872\\
2.78392709181553	1.09528109753684\\
3.16402633461875	1.0958156537792\\
3.59602184827055	1.09605644176465\\
4.08699921102184	1.09615317752435\\
4.64501141975166	1.09618594083084\\
5.27921097499522	1.09618594083084\\
6	1.09618594083084\\
50	1.09618594083084\\
200	1.09618594083084\\
1000	1.09618594083084\\
};
    \end{axis}

\end{tikzpicture}%